\documentclass[11pt]{article}
\usepackage{amsmath,amsfonts}
\def \beq  {\begin{equation}}
\def \eeq  {\end{equation}}
\def \beqar {\begin{eqnarray}}
\def \eeqar {\end{eqnarray}}
\def\sqr#1#2{{\vcenter{\vbox{\hrule height.#2pt
\hbox{\vrule width.#2pt height#1pt \kern#1pt
\vrule width.#2pt}\hrule height.#2pt}}}}

\def\L {{\cal L}}

\def\vx {{\vec x}}
\def\vX {{\vec X}}

\def\vj {{\vec j}}

\def\vo {{\vec \omega}}

\def\pr {\prime}
\def\del {\partial}

\def\a {\alpha}
\def\b {\beta}
\def\g {\gamma}
\def\e {\epsilon}
\def\ve {\varepsilon}
\def\d {\delta}
\def\r {\rho}

\def\l {\lambda}
\def\o {\omega}
\def\Om {\Omega}

\def\H {{\cal H}}

\def\E {{\cal E}}

\def \L {{\cal L}}
\def \H {{\cal H}}

\begin{document}
%%%%%%%%%%%%%%%%%%%%%%%%%%%%%%%%%%%%%%%%%%%%%%%
%\fontfamily{pnb}\fontsize{12pt}{16pt}\selectfont
%\fontfamily{pzc}\fontsize{14pt}{16pt}\selectfont
%\fontfamily{pbk}\fontsize{12pt}{16pt}\selectfont
%\fontfamily{cmr}\fontsize{11pt}{16pt}\selectfont
%\fontfamily{phv}\fontshape{ro}\fontsize{11pt}{14pt}\selectfont
%\fontfamily{ptm}\fontseries{m}\fontshape{r}\fontsize{12pt}{16pt}\selectfont
%\fontfamily{pnc}\fontseries{m}\fontshape{r}\fontsize{11pt}{15pt}\selectfont
%\usefont{T1}{phv}{m}{it}
%%%%%%%%%%%%%%%%%%%%%%%%%%%%%%%%%%%%%%%%%%%%%%%
\def \CMP {{ Commun. Math. Phys.}}
\def \PRL {{ Phys. Rev. Lett.}}
\def \PL {{Phys. Lett.}}
\def \NPBProc {{ Nucl. Phys. B (Proc. Suppl.)}}
\def \NP {{ Nucl. Phys.}}
\def \RMP {{ Rev. Mod. Phys.}}
\def \JGP {{ J. Geom. Phys.}}
\def \CQG {{ Class. Quant. Grav.}}
\def \MPL {{Mod. Phys. Lett.}}
\def \IJMP {{ Int. J. Mod. Phys.}}
\def \JHEP {{ JHEP}}
\def \PR {{Phys. Rev.}}
\def \JMP {{J. Math. Phys.}}
\def\JoP {{J. Phys.}}
%%%%%%%%%%%%%%%%%%%%%%%%%%%%%%%%%%%%%%%%%%%%%%%
%%%%%%%%%%%%%%%%%%%%%%%%%%%%%%%%%%%%%%%%%%%%%%%
\begin{titlepage}
\null\vspace{-62pt} \pagestyle{empty}
\begin{center}
\rightline{CCNY-HEP-06/x} \rightline{December 2006} \vspace{1truein}
{\Large\bfseries

Stability and Evolution of Color Skyrmions in the Quark-Gluon Plasma
}\\
\vskip .2in\noindent

%%%%%%%%%%%%%%%%%%%%%%%%%%%%%%%%%%%%%%%%%%%%%%%%%
\vspace{.5in}
{\bf\large JIAN DAI}\footnote{E-mail: \fontfamily{cmtt}\fontsize{11pt}{15pt}\selectfont
jdai@sci.ccny.cuny.edu}
% and {\bf\large V. P. NAIR}\footnote{E-mail: \fontfamily{cmtt}\fontsize{11pt}{15pt}\selectfont vpn@sci.ccny.cuny.edu}
\\
\vspace{.15in}{\itshape Physics Department\\
City College of the CUNY\\
New York, NY 10031}\\

\fontfamily{cmr}\fontsize{11pt}{15pt}\selectfont
 \vskip 1in
%%%%%%%%%%%%%%%%%%%%%%%%%%%%%%%%%%%%%%%%%%%%%%%%%%%%%%%%%%%%
\centerline{\large\bf Abstract}
\end{center}

We show the existence of unstable color skyrmions in a class of
nonabelian fluid models. Oscillating and expanding solutions are
found in the time-dependent case.

\end{titlepage}
%%%%%%%%%%%%%%%%%%%%%%%%%%%%%%%%%%%%%%%%%%%%%%%%%%%%%%
\pagestyle{plain} \setcounter{page}{2}
\setcounter{footnote}{0}

\section{Introduction}

 Hydrodynamics is the natural framework for the nuclear matter at finite temperature and at long distance scales.
 A novel nuclear matter state has been discovered in the Relativistic Heavy Ion Collision (RHIC) experiment \cite{expt1}
 (for review see for example \cite{expt2}). It is widely believed that a deconfinement phase transition happens
 and the corresponding deconfined state is the {\em quark gluon plasma}
 (QGP) \cite{Shuryak} (for early work prior to QCD, see \cite{Lee}).
 It has been estimated that the critical density and critical temperature
 are $\mu_C\sim 1 GeV/fm^3$, ten times the density of the normal nuclear matter, and
 $T_C\sim 170 MeV$ at which $\alpha_s\sim 1$. So the reliability of the perturbative description of
 QGP is highly suspicious. To apply hydrodynamics in this case
 requires the consideration of the propagation of the color charges.
 Therefore the corresponding liquid is ``colored''. More interestingly,
 experiment shows that the shear viscosity of the deconfined nuclear matter is close to the lower
 bound set by theory based on AdS/CFT \cite{grav}. This implies that a perfect
 fluid is a good approximation to the real QGP. The framework for a perfect color liquid
 has been developed in \cite{NFM} (for review see \cite{NFMR}).

 In using the nonabelian fluid mechanics to explore the physics of
 QGP, the natural first step is to discuss the classical solution
 space. So we ask whether there exists any topological nontrivial
 solution. In fact within the framework of the nonabelian fluid,
 the configuration of fluid is described by a field $g$ which takes
 values
 in the color group, say $SU(n)$. Because QGP only exists in finite space region,
 a natural boundary condition is that $g$ should go to a constant at spacial infinity. Therefore, the fluid
 configurations are classified by $\Pi_3(SU(n))=\mathbb{Z}$.
 Moreover, in each topological class, we hope to check whether there
 exists a configuration minimizing the total energy. Such solutions are referred to as
 skyrmions related to color group or simply {\em color skyrmions}. In our previous
 paper \cite{DN}, the existence of color skyrmion is shown for a particular choice of the hydrodynamic Hamiltonian.
 In this paper, we continue the discussion of the existence of color
 skyrmions in both time-independent and time-dependent cases.

 We will consider the nonabelian fluid system whose equation of state (EOS) follows the so-called
 {\em $\g$-law}, namely the pressure density is
 proportional to the energy density:
 \beq\label{GammaLaw}
  \wp=(\g-1)\ve.
 \eeq
 From the relativistic fluid mechanics,
 we know that the case $\g=4/3$ corresponds to radiation
 ($\wp=\ve/3$) while the case $\g=1$ corresponds to dust ($\wp=0$).
 The up-to-date results on EOS for the isentropic expansion of QGP come from
 lattice simulation \cite{lat}. In the temperature region relevant to RHIC, for example,
 a phenomenological parametrization of the resulting equation of state is given by
 \beq\label{EOS}
  {\wp\over\ve}={1\over
  3}\Bigl(1-\frac{1.2}{1+0.5\ve~\mbox{fm}^3/\mbox{GeV}}\Bigr).
 \eeq
 We see it is a smooth transition from radiation to dust.
 In addition, we will take $\g=2$ as a heuristic example.
 Consequently, we will restrict $1\le \g\le 2$ in this
 paper.
 Our major results can be summarized as follows. For the
 time-independent case,
 \begin{itemize}
  \item $\g=2$: we showed the existence of color skyrmion for this
  case in a particular ansatz for configuration in previous work. In
  this paper, the existence is shown for more general cases.
  \item $6/5\le\g\le 5/3$: the topological configuration are in general unstable.
  The total energy gets minimized by infinite dilution in the fluid and the final value of energy is zero.
  \item $\g=1$: the energy is minimized to a finite value when the
  fluid is infinitely diluted.
  \item $1<\g<6/5$ and $5/3<\g<2$: unstable topological configurations exist; however,
  the existence of any metastable soliton cannot be ruled out.
 \end{itemize}
 For time-dependent case, both oscillating and expanding solutions are
 found.

 The organization of this paper is the following.
 In Sect.~\ref{Pre}, preliminaries materials on nonabelian fluid
 mechanics are reviewed.
 In Sect.~\ref{TIS} and \ref{DG}, static color skyrmions are thoroughly considered for different $\g$.
 In Sect.~\ref{TDS}, time-dependent solutions are given.
 In Sect.~\ref{DIS}, conclusions are given and some open issues are addressed.

\section{Preliminaries}\label{Pre}

 We describe the internal degrees of freedom for a single colored particle by a group element $g$ in the color group
 $SU(n)$. And the corresponding phase space structure is determined by the
 symplectic potential:
 \beq
  \Theta=-i\r_sTr(T_sg^{-1}dg)
 \eeq
 from which a key Poisson bracket is determined
 \beq
  \{\r_s, g\}=-igT_s.
 \eeq
 $T_s$ is a generator in Cartan subalgebra and $\r_s$ is the
 conjugate momentum. The index $s$ runs over all the generators
 $T_s$. The dynamics of color degrees of freedom is therefore
 determined by the Lagrangian
 \beq
  L=-i\r_sTr(T_sg^{-1}\dot{g})-H(\r_s,g)
 \eeq
 where $H$ is the Hamiltonian.

 The generalization to many-particle system gives the nonabelian fluid
 Lagrangian density:
 \beq\label{A1}
  \L=\sum\limits_sj_s^\mu \o_{s\mu}-F(\{n_s\})
 \eeq
 where $j_s^\mu$ is the color current,
 $n_s^2=j_s^\mu j_{s\mu}$ and $F$ an invariant potential function.
 And
 \beq
  \o_{s\mu}=-iTr(T_sg^{-1}\del_\mu g)
 \eeq
 is the generalized velocity field. The action in a general metric background is given
 by $S=\int d^4x\sqrt{g}\L$.
 In this paper, we will only consider rank-one group $SU(2)$; therefore, the
 index $s$ will be omitted hereafter.

 Recall that the energy-momentum tensor of an ideal fluid is of the standard form
 \beq\label{A4}
  T^{\mu\nu}=(\ve+\wp)u^\mu u^\nu-\wp g^{\mu\nu}
 \eeq
 where $u^\mu$ is the unit velocity field such that $u^\mu u_\mu=1$.
 Now we identify $u^\mu$ in the following way:
 \beq\label{A5}
  j^\mu=nu^\mu.
 \eeq
 Note an on-shell condition is
 \beq\label{A6}
  n\o_\mu=F^\pr j_\mu.
 \eeq
 With the identity $\d g = gg^{\mu\nu}\d g_{\mu\nu}$ and the definition of the energy-momentum
 tensor,
 \beq\label{A3}
  T^{\mu\nu}=-{2\over\sqrt{g}}{\d S\over\d g_{\mu\nu}},
 \eeq
 it can be verified that $T^{\mu\nu}$ for the nonabelian fluid system in Eq.~(\ref{A1}) is in
 accordance with the form for an ideal fluid in Eq.~(\ref{A4}) and that
 \beq\label{ID}
  \ve=F,~~\wp=nF^\pr-F.
 \eeq
 So we claim that the Lagrangian in Eq.~(\ref{A1}) does describe an
 ideal fluid system.
 Eq.~(\ref{ID}) shows that for the $\g$-law of Eq.~(\ref{GammaLaw}), we solve
 \beq\label{A8}
  F(n)={\a\over\g} n^\g,
 \eeq
 with $\a$ a dimensional constant.

 Remember the physical boundary condition for $g$ as a field defined
 in space $\mathbb{R}^3$ in the previous introductory section. In accordance with the boundary condition,
 the configuration space is classified by the mapping class $g:S^3\rightarrow
 SU(n)$. We do not need to specify the topological number in this
 paper. The assumption that the configuration carries certain
 topological charges is enough in practice below.

\section{Variational Analysis of Time-Independent Configurations}\label{TIS}

 We discuss the existence of the static soliton solution with defined topological charge in the nonabelian fluid
 in this section.

 \subsection{General Setting}

 To do so, it is convenient to use Hamiltonian. In
 fact, the Lagrangian in Eq.~(\ref{A1}) can be reexpressed as
 $\L=\r\o_0-\H$ where $\r=j^0$. The Hamiltonian density is given by
 \beq\label{A10}
  \H=\vj\cdot\vo+F
 \eeq
 and there is an on-shell constraint:
 \beq
  {\d \H\over \d \vec{j}}=\vec{\o}+{\d
  F\over\d\vec{j}}=\vec{\o}-{F^\prime\over n}\vec{j}=0.
 \eeq
 With some simple algebra, we get a dynamical system
 \beq
  \H={n\over F^\prime}\o^2+F, ~
  \r^2=n^2+{n^2\over F^{\prime 2}}\o^2
 \eeq
 where $\o=|\vo|$. In principle, the invariant density $n$ can be eliminated from Hamiltonian by
 solving the second equation and the Hamiltonian becomes a function of
 $\r$ and $\o$. However, the constraint is in general unsolvable
 algebraically. This is the hard part of this issue.

 Now we introduce the major physical assumptions of this paper,
 under which the existence of color skyrmions are considered.
 \begin{enumerate}
  \item We only consider the fluid configurations with a single characteristic scale, which is written as
  $R$.
  \item $\r$ scales like $1/R^3$.
  \item $\o$ scales like $1/R$.
  \item $F$ follows the $\g$-law.
 \end{enumerate}
 \noindent
 We will simply call $R$ as the size of fluid configuration.
 Remarkably, these assumptions are general enough regardless of the spin and
 the color of the fluid configuration.
 Accordingly, we introduce the dimensionless quantities in the following:
 \beq
  \vX=\vx/R,~~
  \mathbf{P}=R^3\r,~~
  \Om=R\o.
 \eeq
 Moreover, we parameterize the dimensional constant in the
 definition of $\g$-law as
 \beq
  \mu=\a^{1/(4-3\g)}.
 \eeq
 So we have two scales in the problem: $\mu$ which is fixed, and $R$
 which can be varied to minimize the energy. It is natural to make the size
 of the configuration dimensionless by defining
 \beq
  r=\mu R.
 \eeq
 Instead of using the invariant density $n$, it is more
 convenient to use the ratio
 \beq
  Z={\r\over n}.
 \eeq
 With these technical preparations and by putting the $\g$-law in, we can recast the physical quantities
 in a more compact form:
 \beqar\nonumber
  H&=&T+P,\\
  \nonumber
  T&=&{\mu\over r^{5-3\g}}\int d^3X{A\over Z^{2-\g}},\\
  P&=&{\mu\over r^{3(\g-1)}}\int d^3X{B\over
  Z^\g};\label{E}\\
  \label{A7Z}
  Z^2&=&1+cZ^{2(\g-1)}.
 \eeqar
 where we introduce the following definitions
 \beq\label{C}
  c={C\over r^{2(4-3\g)}}
 \eeq
 and
 \beq\label{PARA}
  A=\Om^2\mathbf{P}^{2-\g},~
  B={\mathbf{P}^\g\over \g},~
  C={\Om^2\over \mathbf{P}^{2(\g-1)}}.
 \eeq
 We will refer $T$ as \underline{tension term} and $P$ as \underline{potential term}.
 We can now define the basic problem of the variational procedure:
 Given $\g$, whether the total energy in Eq.~(\ref{E}) is able to be minimized by the variation of $r$ with
 the constraint given in Eq.~(\ref{A7Z})? The solution to this question gives rise to a stable configuration
 in nonabelian fluid within the variational method.

 To solve this question, we {\em first} consider some general implications
 of the constraint equation (\ref{A7Z}) and the solution $Z$ as a function in $c$.
 Eq.~(\ref{A7Z}) cannot be given algebraically for the generic value of $\g$. In other words,
 Eq.~(\ref{A7Z}) is transcendental in general; even for the simple rational powers like
 $\g=7/6,6/5,7/5,8/5,9/5,11/6$, the roots of this equation cannot be
 written down. Now we rewrite Eq.~(\ref{A7Z}) in another form
 \beq\label{A7c}
  c={Z^2-1\over Z^{2(\g-1)}}.
 \eeq
 we see immediately that $Z>1$ for all $c>0$ since $Z$ as a ratio of two densities is always nonnegative.
 \footnote{Technically, $\r$ can be zero at some points. However,
 such points have no contribution to energy; therefore in the
 integral they can be excluded.}
 The next step is to consider the monotonicity of $Z$ in $c$.
 We take derivatives on both sides of Eq.~(\ref{A7c}) to get
 \beq
  {dc\over dZ}={2(2-\g)\over Z^{2\g-3}}+{2(\g-1)\over Z^{2\g-1}}.
 \eeq
 So as long as $1<\g<2$, $dc/dZ>0$ and $Z$ is strictly monotonic in $c$.
 Henceforth, we will assume
 \beq
  1\le \g\le 2.
 \eeq
 Furthermore, we claim a very useful inequality:
 \beq\label{Ie}
  Z>(\g-1)cZ^{2\g-3}.
 \eeq
 In fact, take derivatives on the both
 side of Eq.~(\ref{A7Z}) and one can derive the following relation:
 \beq
  Z-(\g-1)cZ^{2\g-3}={Z^{2\g-2}\over 2Z^\prime}.
 \eeq
 The claimed statement follows the fact that the righthand side of
 above equation is positive. So much for our general consideration
 of $Z$.

 \subsection{$\g=4/3$}
 In the following we will discuss the existence of variational
 solution for different $\g$.
 Recall the definition $c=Cr^{2(3\g-4)}$. An important
 observation is that $\g=4/3$ is a critical value for which the
 density ratio $Z$ is independent of the configuration size $R$.
 So even though the constraint in Eq.~(\ref{A7Z}) can be solved algebraically in this case,
 the detail of the solution is irrelevant to the existence of stable
 configuration. In fact, the total energy is of the form
 \beq
  E={const.\over R}
 \eeq
 where the unspecified constant comes from the dimensionless
 integral. Physically, this simplicity follows directly from the
 fact that the case $\g=4/3$ describes the massless particle system.
 The immediate inference is that no stable fluid
 configuration with fixed topological charge exists at finite $R$.
 For the physics of QGP, this is not bad news because we
 expect to see the expansion process of the fireball of nuclear
 matter. So we refer this type of unstable configuration as the {\em
 unstable color skyrmion}. We can imagine that once an unstable configuration is generated,
 it can be stabilized through a process
 in which the size $R$ is enlarged and the energy is dissipated. Because the process is continuous,
 the topological charge is not changed during this process.
 However, the final state is a null state in which the configuration is completely diluted and the energy
 is completely dissipated.

 \subsection{Asymptotic Analysis}

 We discuss the asymptotic behaviors of the total energy in
 the limits $r\rightarrow 0$ and $\infty$ for $1<\g<4/3$ and $4/3<\g<2$. To do so, we need to
 solve Eq.~(\ref{A7Z}) approximately in these two limits.
 In fact if $c=0$ the physical solution to Eq.~(\ref{A7Z}) is $Z=1$.
 So, for small $c$, the equation can be solved approximately by
 \beq
  Z\approx 1+{c\over 2}.
 \eeq
 For large $c$, we recast Eq.~(\ref{A7Z}) in the form
 \beq\label{Z}
  {Z^2\over c}={1\over c}+Z^{2(\g-1)}.
 \eeq
 Let $Z\sim c^\l$ with a positive power $\l$ such that the first term on the right hand
 side of (\ref{Z}) can be dropped. The approximative solution is
 \beq
  Z=c^{1/2(2-\g)}.
 \eeq

 In the region $4/3<\g<2$ for small $r$, the contribution of
 $Z$ can be omitted and the potential term is the dominant part of
 the energy in Eq.~(\ref{E}),
 \beq
  H\sim {\mu\over r^{3(\g-1)}}.
 \eeq
 In the region $4/3<\g<2$ for large $r$,
 \beq\label{ASYMPT}
  Z\sim r^{(3\g-4)/(2-\g)},~~
  T\sim {\mu\over r},~~
  P\sim {\mu\over r^{(5\g-6)/(2-\g)}};
 \eeq
 therefore, the tension term dominates.
 In the region $1<\g<4/3$ for large $r$, the contribution of
 $Z$ can be omitted and again the potential term is dominant.
 In the region $1<\g<4/3$ for small $r$ from the same equation
 (\ref{ASYMPT}), the tension term dominates. The common
 feature of both regions of $\g$ is that the energy blows up in the
 $r\rightarrow 0$ limit and vanishes as $r\rightarrow\infty$. So the
 most immediate conclusion from above asymptotic analysis is that
 there always exists unstable color skyrmion in each region for $\g$.

 Nevertheless, there is a subtle sub-region for both $1<\g<4/3$ and
 $4/3<\g<2$ from the behavior of energy in the small
 $r$ limit. In the sub-region $5/3<\g<2$ for small $r$, the
 tension term behaves like
 \beq\label{ASYM1}
  T\sim \mu r^{3\g-5}\rightarrow 0.
 \eeq
 Combining the second formula in Eq.~(\ref{ASYMPT}) and
 Eq.~(\ref{ASYM1}), we conclude the profile of tension term has
 maximum for some finite $r$ in the region $5/3<\g<2$.
 In the sub-region $1<\g<6/5$ for small $r$, the
 potential term behaves like
 \beq\label{ASYM2}
  P\sim \mu r^{(6-5\g)/(2-\g)},
 \eeq
 which implies that the profile of potential term has
 maximum for some finite $r$ in the region $1<\g<6/5$.
 These results
 reveal the possibility that there may exist
 metastable color skyrmions in these two sub-regions. By a metastable
 soliton, we mean that the energy has a local minimum. The implication for physics
 is the following.
 At the classical level, if the configuration is
 generated with a large enough size $R$ then it is unstable and
 it can only be stabilized by expansion and dissipation
 until the configuration is completely diluted.
 For
 small sized configuration however, there is a locally stable point
 with finite energy.

 \subsection{$6/5<\g<5/3$}

 Beyond the asymptotic analysis, we conduct a monotonicity analysis to scrutinize
 a claim that {\em there does not exist metastable color skyrmion in
 $6/5<\g<5/3$}.
 This conclusion is certainly true for $\g=4/3$ from our previous discussion.
 For $4/3<\g<5/3$, the conclusion is also easy to make. In fact,
 from the monotonicity of $Z$ in $c$ and the dependency of $c$ on $r$, it is easy to conclude
 that the total energy is strictly decreasing in $r$. And the two limiting cases $r\rightarrow 0, \infty$
 have been considered in the asymptotic analysis.

 The case for $6/5<\g<4/3$ is more subtle. The monotonicity of total energy is
 determined by the two combinations,
 \beq\label{W1}
  W_1={Z\over c^{3(\g-1)\over 2\g(4-3\g)}},~~
  W_2={Z\over c^{5-3\g\over 2(4-3\g)(2-\g)}},
 \eeq
 which are implicit in the expressions of potential term and tension term in total energy.
 By Eq.~(\ref{A7Z}), we know $W_2$ satisfies the relation:
 \beq
  W_2^2=c^{-{3\g-5\over (3\g-4)(2-\g)}}+c^{1\over 3\g-4} W_2^{2(\g-1)}.
 \eeq
 Taking derivatives on both sides,
 \beq\label{D}
  2\Bigl(W_2-{\g-1\over c^{1\over 4-3\g}} W_2^{2\g-3}\Bigr){dW_2\over dc}=(-)
  \frac{Z^{2(\g-1)}+{5-3\g\over (2-\g)c}}{(4-3\g)c^{{5-3\g\over
  (4-3\g)(2-\g)}}},
 \eeq
 where we have expressed the right hand side back in terms of $Z$.
 The bracketed factor on the left hand side can be also expressed in
 terms of $Z$ as
 \beq
  W_2-{\g-1\over c^{1\over 4-3\g}} W_2^{2\g-3}={Z-(\g-1)cZ^{2\g-3}\over
  c^{(3\g-5)\over 2(3\g-4)(2-\g)}}>0.
 \eeq
 The last inequality is because of (\ref{Ie}) so we conclude that
 \beq
  {dW_2\over dc}<0,
 \eeq
 namely $W_2$ is strictly decreasing in $c$, hence increasing in
 $r$.
 Due to Eq.~(\ref{A7Z}), $W_1$ satisfies
 \beq
  W_1^2=c^{-{3(\g-1)\over\g(4-3\g)}}+c^{6-5\g\over\g(4-3\g)}W_1^{2\g-2}.
 \eeq
 Taking derivatives on both sides of Eq.~(\ref{W1}),
 \beq\label{W1I}
  2(W_1-{\g-1\over c^{5\g-6\over \g(4-3\g)}}W_1^{2\g-3}){dW_1\over
  dc}=(-)\frac{(5\g-6)Z^{2(\g-1)}+{3(\g-1)\over c}}{\g(4-3\g)c^{3(\g-1)\over
  \g(4-3\g)}}.
 \eeq
 Again, the sign of $dW_1/dc$ is determined by the righthand side
 because the inequality (\ref{Ie}) implies the left hand side is positive.
 So Eq.~(\ref{W1I}) implies
 \beq
  {dW_1\over dc}<0
 \eeq
 namely $W_1$ is also strictly decreasing in $c$.
 \footnote{Certain uniformity conditions for the profile functions $\Om$ and $\mathbf{P}$ are
 expected to guarantee that the conclusion does not change after the integral over the space coordinates $\int d^3X$.}
 To conclude this part of the monotonicity analysis, we see the total
 energy is strictly decreasing in $r$ in the region $6/5<\g<5/3$.
 In addition, the monotonicity cannot be established from the
 righthand sides of Eqs.~(\ref{D},\ref{W1I}) if $\g>5/3$ or
 $\g<6/5$. We leave the issue for possible metastable configurations to further work.

\subsection{$\g=2$}

 Now we consider the boundary value of $\g=2$.
 This is the case which has been worked out with a more specific ansatz in our
 previous paper \cite{DN}. The same result can be established on more general grounds here.
 For $\g=2$, Eq.~(\ref{A7Z}) can be solved by
 \beq
  n=\sqrt{\r^2-{\o^2\over\a^2}}.
 \eeq
 The Hamiltonian density in Eq.~(\ref{A10}) is then
 \beq
  \H={\o^2\over 2\a}+{\a\r^2\over 2}.
 \eeq
 The total energy is given by
 \beq
  H={\mu\over 2}\int d^3X \Bigl(r\Om^2+{\mathbf{P}^2\over r^3}\Bigr).
 \eeq
 The existence of color skyrmion that minimizes the energy at certain finite $R$
 is straightforward.

\subsection{$\g=1$}

 In this case, Eq.~(\ref{A7Z}) is solved by
 \beq
  n={\r\over\sqrt{1+{\o^2\over \a^2}}}
 \eeq
 and the energy is
 \beq
  E=\mu\int d^3X ~\mathbf{P}\sqrt{1+{\Om^2\over r^2}}.
 \eeq
 The unstable color skyrmion has a finite energy in the limit $R\rightarrow \infty$
 and the final state energy is
 \beq
  E\longrightarrow \mu\int d^3X ~\mathbf{P}.
 \eeq

 To this point, we establish the results of the time-independent color skyrmions listed in the introductory
 section.

\section{$\g$-Dependence of Total Energy}
\label{DG}
 In previous section, we solve the variational problem for the
 nonabelian fluid system with fixed $\g$. However, in reality the
 power $\g$ is a changing quantity during the process of expansion
 of QGP. So in this section, we consider the dependence of the total energy on the
 power $\g$.
 To do so, we will calculate $\del H/\del\g$.
 We assume the profiles of $\Om$, $\mathbf{P}$ and the parameters $\mu$,
 $R$ are independent to $\g$.
 We claim that $\del H/\del\g<0$ in the ``physical parameter region''
 $1<\g<4/3$ provided the soliton size $R$ is large enough.
 In fact,
 \beqar\nonumber
  {1\over\mu}{\del H\over \del \g}&=&\int d^3X\Bigl(({\Om^2\mathbf{P}^{2-\g}\over r^{5-3\g}Z^{2-\g}}
  -{\mathbf{P}^\g\over \g r^{3(\g-1)}Z^\g})\ln{r^3 Z\over \mathbf{P}}\\\nonumber
  &&-\Bigl[(2-\g){\Om^2\mathbf{P}^{2-\g}\over r^{5-3\g}Z^{3-\g}}
  +{\mathbf{P}^\g\over r^{3(\g-1)}Z^{\g+1}}\Bigr]{\del Z\over\del\g}\\
  &&-{\mathbf{P}^\g\over \g^2r^{3(\g-1)}Z^\g}
  \Bigr).
 \eeqar
 We can impose the following sufficient conditions to guarantee $\del H/\del \g<0$:
 \beqar\label{I1+}
  {Z^{2(\g-1)}\over r^{2(4-3\g)}}&<&{\mathbf{P}^{2(\g-1)}\over \g
  \Om^2};\\\label{I2+}
  {\del Z\over\del\g}&>&0;\\\label{I3+}
  r^3Z&>&\mathbf{P}.
 \eeqar

 By using Eq.~(\ref{A7Z}) and the definition of $C$ in
 Eq.~(\ref{PARA}), one can show that (\ref{I1+}) is equivalent to
 \beq\label{I1+M}
  Z^2<{\g+1\over\g}.
 \eeq
 By the fact that $dZ/dc>0$ and the relation in Eq.~(\ref{C}), we
 conclude that for large enough $r$, the inequality (\ref{I1+M}),
 hence (\ref{I1+}), is satisfied if $1<\g<4/3$.
 For (\ref{I2+}), we take derivatives with respect to $\g$ on both sides of
 (\ref{A7Z}),
 \beq
  \Bigl(Z-(\g-1)cZ^{2\g-3}\Bigr){\del Z\over\del\g}=
  cZ^{2\g-2}\ln{r^3Z\over \mathbf{P}}.
 \eeq
 Recall the relation in (\ref{Ie}). So (\ref{I2+}) is satisfied if (\ref{I3+}) is
 satisfied. To show (\ref{I3+}), we need to check the monotonicity of the
 combination
 \beq
  W_3\equiv c^{3/(6\g-8)}Z.
 \eeq
 We claim that
 \beq\label{CLAIM}
  {dW_3\over dc}<0.
 \eeq
 In fact, $W_3$ satisfies the equation
 \beq
  W_3^2=c^{3/3\g-4}+c^{2/3\g-4}W_3^{2(\g-1)}.
 \eeq
 Take derivatives in $c$:
 \beq\label{D1+}
  2(W_3-(\g-1)c^{2/3\g-4}W_3^{2\g-3}){dW_3\over dc}=-{1\over 4-3\g}(3c^{7-3\g/3\g-4}+2c^{6-3\g/3\g-4}W_3^{2\g-2}).
 \eeq
 It is easy to show
 \beq
  W_3-(\g-1)c^{2/3\g-4}W_3^{2\g-3}=c^{3/6\g-8}(Z-(\g-1)cZ^{2\g-3}).
 \eeq
 Therefore, this factor is positive. So from (\ref{D1+}), it is
 obvious that $dW_3/dc<0$ for $\g<4/3$. We see that the conditions (\ref{I2+}) and (\ref{I3+}) are
 universally true provided $\g$ is in the physical region. But
 (\ref{I1+}) requires a large configuration size.
 %\footnote{Please note again that as a physicist's proof, we don't take care of the
 %uniformity of $\Om$ and $\mathbf{P}$ as well as the (possible) subtleties in the
 %limits $\g\rightarrow 1, 4/3$.}

 Now we want to estimate the lower bound for $r$.
 From (\ref{A7Z}) and (\ref{I3+}), we derive an interesting
 relation:
 \beq
  Z^2>1+{\Om^2\over r^2}.
 \eeq
 Combined with (\ref{I1+M}),we have
 \beq
  1+{\Om^2\over r^2}<Z^2<{1+\g\over \g}\Rightarrow
  r^2>\g\Om^2.
 \eeq
 So given an ansatz profile function $\Om$, we know how to estimate the configuration size
 such that the energy is strictly decreasing with $\g$.

 The physical meaning of the condition $\del H/\del \g<0$
 is the following. It is easy to understand from the previous
 section that with expanding soliton size $r$, the energy decreases.
 On the other hand, the equation of state in Eq.~(\ref{EOS}) implies
 with the decreasing energy density, $\g$ becomes smaller as well.
 So there is a trajectory in the $(\g,r)$-plane along which the
 total energy can be expected to be unchanged during the expansion
 process.

\section{Time Evolution of Color Skyrmions}\label{TDS}

 So far we have dealt with time-independent fluid configurations.
 In this section, we consider the time-dependent configurations.
 To do so, we need to take account of the presence of $\o_0$ which plays no role in the time-independent
 case. Accordingly, we need to eliminate $j^0$ as well as $\vec{j}$ by
 the equation of motion,
 \beq
  {\d \L\over\d j^\mu}=\o_\mu-\a n^{\g-2}j_\mu=0
 \eeq
 where we have used the $\g$-law.
 Eliminating $j^\mu$ by $j^\mu \o_\mu=\a n^\g$, $\o^\mu \o_\mu=\a n^{\g-2}j_\mu\o^\mu$, we get
 \beq
  \o^\mu\o_\mu=\a^2n^{2\g-2}.
 \eeq
 The Lagrangian density is therefore expressed in terms of
 $\o_\mu$ as
 \beq
  \L={\g-1\over\g}\a n^\g =
  {\g-1\over\g}\a^{-{1\over\g-1}}\Bigl(\o^\mu\o_\mu\Bigr)^{\g\over
  2(\g-1)}.
 \eeq
 Since in this paper we will concentrate on classical
 configurations, the pre-factor $(\g-1)/\g\a^{1/\g-1}$ does not matter
 to us at this level. So we will deal with the following Lagrangian
 \beq\label{C3}
  L=\int d^3x \Bigl(\o_0^2-\vec{\o} \cdot\vec{\o}\Bigr)^{\g\over
  2(\g-1)}.
 \eeq

 As in the case for the time-independent configurations, we make the following scaling assumptions:
 \beq
  \o_0^2=f_1{\dot{R}^2\over R^2},~~
  \vec{\o}\cdot\vec{\o}={f_2\over R^2}
 \eeq
 where $f_{1,2}$ are two dimensionless functions depending only on
 $\vec{X}=\vec{x}/R$. Then the Lagrangian in Eq.~(\ref{C3}) is
 transformed to be
 \beq\label{C4}
  L(R,\dot{R})=R^{2\g-3\over\g-1}\int d^3X \Bigl(
  f_1\dot{R}^2-f_2\Bigr)^{\g\over 2(\g-1)}.
 \eeq
 Next we will derive the Euler-Lagrangian equation for
 the Lagrangian in Eq.~(\ref{C4}).
 First of all,
 \beq
  {\del L\over\del \dot{R}}={\g\over\g-1} R^{2\g-3\over \g-1}\dot{R}
  \int d^3X f_1\Bigl(f_1\dot{R}^2-f_2\Bigr)^{2-\g\over 2(\g-1)},
 \eeq
 from which
 \beqar\nonumber
  {d\over dt}{\del L\over\del\dot{R}}&=&{\g\over \g-1}\Biggl(
  {2-\g\over \g-1}R^{2\g-3\over \g-1}\dot{R}^2\ddot{R}\int d^3X
  f_1^2\Bigl( f_1\dot{R}^2-f_2\Bigr)^{4-3\g\over 2(\g-1)}\\\nonumber
  &&
  +R^{2\g-3\over\g-1}\ddot{R}\int d^3X
  f_1\Bigl(f_1\dot{R}^2-f_2\Bigr)^{2-\g\over 2(\g-1)}\\&&
  +{2\g-3\over \g-1} R^{\g-2\over\g-1}\dot{R}^2\int d^3X
  f_1\Bigl(f_1\dot{R}^2-f_2\Bigr)^{2-\g\over 2(\g-1)}
  \Biggr).
 \eeqar
 Further
 \beq
  {\del L\over\del R}={2\g-3\over\g-1} R^{\g-2\over \g-1}
  \int d^3X \Bigl(f_1\dot{R}^2-f_2\Bigr)^{\g\over 2(\g-1)}.
 \eeq
 Thus the Euler-Lagrangian equation is given by
 \beqar\nonumber
  {2-\g\over\g-1}R\dot{R}^2\ddot{R}\int d^3X f_1^2\Bigl( f_1\dot{R}^2-f_2\Bigr)^{4-3\g\over
  2(\g-1)}&&\\\nonumber
  +R\ddot{R}\int d^3X
  f_1\Bigl(f_1\dot{R}^2-f_2\Bigr)^{2-\g\over 2(\g-1)}&&\\\nonumber
  +{2\g-3\over\g-1}\dot{R}^2\int d^3X f_1\Bigl(f_1\dot{R}^2-f_2\Bigr)^{2-\g\over
  2(\g-1)}&&\\
  -{2\g-3\over\g}\int d^3X \Bigl(f_1\dot{R}^2-f_2\Bigr)^{\g\over
  2(\g-1)}&=&0.\label{ELE}
 \eeqar
 In the following we will investigate into two particular cases for
 $\g=2$ and $\g=4/3$.

\subsection{$\g=2$}

 In this case, stable time-independent color skyrmion exists. So we expect
 the solution to the time-evolution equation to be of the form of oscillations around the stable point.
 In fact, the Euler-Lagrangian equation (\ref{ELE}) reduces to
 \beq\label{ELE2}
  R\ddot{R}+{\dot{R}^2\over 2}+\mu=0
 \eeq
 where
 \beq
  \mu=\frac{\int d^3Xf_2}{2\int d^3Xf_1}.
 \eeq
 To solve Eq.~(\ref{ELE2}), we change its form to
 \beq
  {d^2\over dt^2}R^{3/2}+{3\mu\over 2R^{1/2}}=0.
 \eeq
 Let $q=R^{3/2}$ and $p=dq/dt$ then
 \beq
  {p\over\mu}{dp\over dq}+{3\over 2q^{1/3}}=0
 \eeq
 which can be integrated as
 \beq\label{TDS2}
  {p^2\over 2\mu}+{9\over 4}q^{2/3}=\E
 \eeq
 where $\E$ is an integral constant.
 The solution in Eq.~(\ref{TDS2}) means the color skyrmion for
 $\g=2$ forms a one-dimensional Hamiltonian system with the
 potential of the form $q^{2/3}$ and the motion is always bounded and oscillating!

\subsection{$\g=4/3$}

 We know that only unstable color skyrmions exist in the
 time-independent case. Therefore, for the time-dependent case, we expect the
 solution to describe the expansion of the color skrymion with the size
 going from some initial value to infinity.
 In fact,
 the Euler-Lagrangian equation (\ref{ELE}) becomes
 \beq\label{ELE4/3a}
  \dot{R}^2(R\ddot{R}-{\dot{R}^2\over 4})-\b_2(R\ddot{R}-{\dot{R}^2\over 2})+\b_4=0
 \eeq
 where
 \beq
  \b_2=\frac{\int d^3Xf_1f_2}{3\int d^3Xf_1^2},~~
  \b_4=\frac{\int d^3Xf_2^2}{12\int d^3Xf_1^2}.
 \eeq
 We will solve Eq.~(\ref{ELE4/3a}) approximately. To do this, we first
 rewrite this equation by the form
 \beq\label{ELE4/3b}
  {R\ddot{R}\over \dot{R}^2}-{1\over 4}={\b_2\over \dot{R}^2}({R\ddot{R}\over\dot{R}^2}-{1\over
  2})-{\b_4\over \dot{R}^4}
 \eeq
 and make the following scaling assumptions:
 \beq\label{Scaling}
  |\dot{R}|\gg 1,~
  |R\ddot{R}/\dot{R}^2|\sim 1,~
  \b_{2,4}\sim 1.
 \eeq

 To the zero order $R\approx R^{(0)}$, Eq.~(\ref{ELE4/3b}) can be approximated by
 \beq
  {R^{(0)}\ddot{R}^{(0)}\over \dot{R}^{(0)2}}-{1\over 4}=0,
 \eeq
 which is solved by
 \beq\label{EX}
  R^{(0)}=R_0({t\over \tau}+1)^{4/3}
 \eeq
 with $R_0$, $\tau$ two integral constants.
 The scaling rule in Eq.~(\ref{Scaling}) works if the ratio
 \beq
  \e={\tau\over R_0}\ll 1.
 \eeq
 It is very interesting to see that $R^{3/4}$ is linear in
 time. Eq.~(\ref{EX}) describes the expansion of the color
 skyrmion for $\tau>0$ and the contraction or collapse for $\tau<0$.
 Actually, we will only take the former case and
 consider the latter as unphysical. We will refer the motion in
 Eq.~(\ref{EX}) as {\em linear expansion}.

 To check the stability of the above-mentioned expansion motion,
 we do the perturbation to the order of $\e^2$ by writing
 \beq
  R=R^{(0)}+\d R
 \eeq
 and introducing the following quantities:
 \beq
  \d R=R_0\e^2y,~
  x={t\over\tau}+1,~
  f^\prime={df\over dx}.
 \eeq
 The perturbation satisfies the following equation
 \beq\label{4/3P}
  y^{\prime\prime}-{2y^\prime\over 3 x}+{4y\over 9x^2}=-{\b_2\over
  4x^{4/3}}.
 \eeq
 Eq.~(\ref{4/3P}) has the particular solution
 \beq
  y={9\b_2\over 8}x^{2/3}.
 \eeq
 It is easy to see that the linear independent solutions to the
 homogenous equation in (\ref{4/3P}) are $x^{4/3}$, $x^{1/3}$. So the
 general solution is given by
 \beq
  y={9\b_2\over 8}x^{2/3}+Ax^{4/3}+Bx^{1/3}
 \eeq
 where $A$, $B$ are two integration constants.
 By examining the initial conditions, we can fix $A$, $B$. Actually
 we let
 \beq
  R|_{t=0}=R^{(0)}|_{t=0},~~
  \dot{R}|_{t=0}=\dot{R}^{(0)}|_{t=0}
 \eeq
 then
 \beq
  y|_{x=1},~
  y^\prime|_{x=1}=0,
 \eeq
 from which we get
 \beq
  A=-{3\over 8}\b_2,~~
  B=-{3\over 4}\b_2.
 \eeq
 The solution to Eq.(\ref{ELE4/3a}),
 to $\e^2$ order, is then given by
 \beq
  R=R_0\Bigl((1-{3\over 8}\b_2\e^2)x^{4/3}+{9\over 8}\b_2\e^2x^{2/3}-{3\over
  4}\b_2\e^2x^{1/3}\Bigr)
 \eeq
 where $x=1+t/\tau$.
 To see the stability of the expansion, we consider the long time
 behavior. It is easy to see that the leading order contribution in
 the limit $t\rightarrow\infty$ is given by the $x^{4/3}$ term, which is the linear expansion
 as in Eq.~(\ref{EX}).

\section{Conclusion and Discussion}
\label{DIS}

 With the assumption of $\g$-law for the equation of state for a
 nonabelian fluid, we show in this paper the value of $\g$ is
 crucial to the existence and the properties of color skyrmion for time-independent configurations. For
 $\g$ between $6/5$ and $5/3$, there is only an unstable color
 skyrmion. For $\g=2$, the skyrmion is stable. The case $\g=1$ is
 special for the unstable skyrmion, since it has finite energy even after it
 is diluted infinitely. For $1<\g<6/5$ and $5/3<\g<2$, we cannot
 rule out the possibility of the existence of metastable skymions
 besides the unstable ones. For two particular values $\g=4/3,2$, we
 furthermore consider the time-dependent configurations. And for the
 latter we find the oscillating evolution and for the former we find
 linear expansion.

 We now turn to the question of how this relates to QGP. As
 mentioned in the introduction, if we try to model the lattice
 estimate of the equation of state as a $\g$-law, the value of $\g$
 varies between $4/3$ and $1$, as the energy density decreases.
 Therefore, our analysis of the expanding soliton for $\g=4/3$
 takes on a special significance. In the creation of the QGP by
 nuclear collision, we are starting at high energy densities, or at
 values of $\g$ close to $4/3$. The collision process also favors
 the creation of small solitons during the phase transition process
 since establishing coherent fields or color densities over large
 scales is generally more difficult, less probable. Based on these
 two observations and in the light of our general analysis, an
 expanding soliton would seem to be the generic case we can expect
 for the QGP. There are, of course, some caveats to our analysis. We
 have neglected thermal gradients as well as dissipative effects
 such as viscosity. (The latter quantity, although small, is not
 zero.) It would be interesting to incorporate some of these
 effects. The soliton by virtue of its nontrivial topology is a
 generic feature of nonabelian fluid dynamics. We may expect many
 qualitative aspects to hold true even with thermal gradients,
 viscosity, etc., although there can be significant changes in
 details.

 We note that the possibility of color skyrmions arises also in the
 recent effective action description of Wilson lines
 \cite{Pisarski1,Pisarski2}. It is clear that skyrmions in the QGP merit
 further analysis. Other hydrodynamics approach towards QGP can be
 found for example in \cite{MM}.

\vskip .2in\noindent

This work was supported in part by a CUNY Collaborative Research
Incentive grant. The author appreciates V. P. Nair's mentoring on
this project.

\end{document}